\documentclass[aip,amsmath,amssymb,twocolumn,]{revtex4}                

\usepackage{textcase}
\usepackage{graphicx}
\usepackage{dcolumn}
\usepackage{bm}
\usepackage{amsmath,amsthm,amssymb,mathrsfs}

\begin{document}

\title{Stable oscillation in spin torque oscillator excited by a small in-plane magnetic field}

\author{Tomohiro Taniguchi${}^{1}$}
\author{Takahiro Ito${}^{2}$}
\author{Yasuhiro Utsumi${}^{2}$}
\author{Sumito Tsunegi${}^{1}$}
\author{Hitoshi Kubota${}^{1}$}

\affiliation{ 
  ${}^{1}$
  National Institute of Advanced Industrial Science and Technology (AIST), Spintronics Research Center, Tsukuba 305-8568, Japan, \\
  ${}^{2}$Faculty of Engineering, Mie University, Tsu, Mi-e, 514-8507, Japan. \\
}

\date{\today}%

\begin{abstract}
Theoretical conditions to excite self-oscillation in a spin torque oscillator 
consisting of a perpendicularly magnetized free layer and an in-plane magnetized pinned layer are investigated 
by analytically solving the Landau-Lifshitz-Gilbert equation. 
The analytical relation between the current and oscillation frequency is derived. 
It is found that a large amplitude oscillation can be excited by applying a small field pointing to the direction 
anti-parallel to the magnetization of the pinned layer. 
The validity of the analytical results is confirmed 
by comparing with numerical simulation, 
showing good agreement especially in a low current region. 
\end{abstract}

\maketitle


\section{Introduction}
\label{sec:Introduction}

High-frequency devices are in demand for practical applications 
such as communication devices, magnetic sensors, and measurement systems. 
Spin torque oscillator (STO) \cite{slonczewski96,berger96,kiselev03,rippard04,houssameddine07,pribiag07,mistral08,kudo09,dussaux10,zeng12,kubota13,lebrun14} 
is a promising candidate for such applications 
because high-frequency signal on the order of Gigahertz can be achieved 
through the giant magnetoresistance (GMR) or tunnel magnetoresistance (TMR) effect. 
A small size on the order of nano-meter, 
a wide frequency tunability over than one Gigahertz, 
and a narrow linewidth less than one Megahertz are also preferable features for device applications. 
In particular, an STO consisting of 
a perpendicularly magnetized free layer \cite{yakata09,ikeda10,kubota12} 
and an in-plane magnetized pinned layer has attracted much attention recently \cite{kubota13} 
because the large relative angle of the magnetizations of the free and pinned layers results in a large emission power 
on the order of 1 $\mu$W \cite{kubota13}. 
Therefore, this type of STO is considered to be the model structure for practical applications. 


The self-oscillation of the magnetization in an STO is excited when several conditions, 
such as the direction and magnitude of an applied field, are satisfied 
\cite{bertotti09book,ebels08,silva10,gusakova11,khalsa15,taniguchi13,taniguchi14APEX,pinna14,taniguchi15}. 
While general conditions to excite the self-oscillation had been formulated \cite{bertotti09book,slavin09,kim12}, 
the possibility to excite the self-oscillation has to be investigated for each problems 
by solving the Landau-Lifshitz-Gilbert (LLG) equation. 
The number of theoretical works studying the conditions of the self-oscillation is few 
because of the complexity arising from the nonlinearity of the LLG equation. 
For example, the oscillation properties of STO 
with the perpendicularly magnetized free layer and the in-plane magnetized pinned layer 
were theoretically studied only in the presence of a perpendicular field \cite{taniguchi13}, 
in which the axial symmetry of the system reduces the difficulty to solve the LLG equation. 
On the other hand, it is still unclear whether the self-oscillation can be excited 
when the field direction is changed. 
It is practically important to investigate 
the possibility to excite self-oscillation 
in the presence of a field in a different direction due to the following reasons. 
For example in the sensor applications, 
an external magnetic field to be detected points to an arbitrary direction. 
Thus, it is insufficient to study the oscillation properties in the presence of a perpendicular field only. 
Another reason relates to the low power consumption. 
Our previous work \cite{taniguchi13} showed that 
the zero-field oscillation is impossible in this type of STO, 
and a relatively high field is necessary to stabilize the self-oscillation by a perpendicular field. 
Discovering a method to excite a stable oscillation by a small field is therefore attractive 
from the viewpoint of reducing the power consumption of the STO. 
The change of the field direction might open a new possibility to this problem. 


In this paper, 
we investigate the oscillation properties of the STO 
in the presence of an in-plane magnetic field 
by solving the LLG equation analytically. 
We find that an applied field smaller than Eq. (\ref{eq:critical_field}) pointing in the direction 
anti-parallel to the magnetization of the pinned layer 
enables to excite a stable self-oscillation. 
This is a preferable feature for practical applications, 
compared with the oscillation by a perpendicular field 
in which relatively large field is required to stabilize the oscillation \cite{taniguchi13}. 
The relation between the current density and the oscillation frequency is derived. 
The comparison with numerical simulation guarantees 
the validity of the theory in the low current region. 


The paper is organized as follows. 
In Sec. \ref{sec:Analytical Solution of LLG Equation}, 
we derived the analytical formulas of 
the work done by spin torque and the dissipation due to damping, 
which are necessary to discuss the possibility to excite self-oscillation, 
by solving the LLG equation. 
Using these results, 
we discuss the theoretical conditions to excite the self-oscillation 
by an in-plane magnetic field 
in Sec. \ref{sec:Analytical Prediction on Possibility to Excite Self-Oscillation}
These analytical discussions are compared with the numerical simulation 
in Sec. \ref{sec:Numerical Simulation}
Section \ref{sec:Discussion} is devoted to the discussion. 
Section \ref{sec:Conclusion} provides our conclusion. 



\begin{figure}
\centerline{\includegraphics[width=0.5\columnwidth]{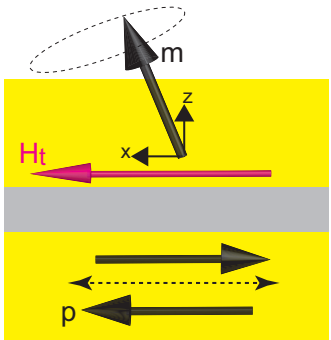}}
\caption{
         Schematic views of STO under consideration. 
         An applied magnetic field points to the positive $x$-direction, 
         whereas the magnetization of the pinned layer is parallel ($\mathbf{p}=+\mathbf{e}_{x}$) 
         or anti-parallel ($\mathbf{p}=-\mathbf{e}_{x}$) to the applied field. 
         \vspace{-3ex}}
\label{fig:fig1}
\end{figure}



\section{Analytical Solution of LLG Equation}
\label{sec:Analytical Solution of LLG Equation}

The system we consider is schematically shown in Fig. \ref{fig:fig1}. 
The unit vectors pointing in the magnetization directions 
of the free and pinned layers are denoted as $\mathbf{m}$ and $\mathbf{p}$. 
We assume that the applied magnetic field points to the positive $x$-direction, 
whereas $\mathbf{p}$ is parallel or anti-parallel to the field. 
We use the macrospin model in this paper because 
its accuracy was confirmed by recent experiment \cite{kubota13}. 
The LLG equation is given by 
\begin{equation}
  \frac{d \mathbf{m}}{dt}
  =
  -\gamma
  \mathbf{m}
  \times
  \mathbf{H}
  -
  \gamma 
  H_{\rm s}
  \mathbf{m}
  \times
  \left(
    \mathbf{p}
    \times
    \mathbf{m}
  \right)
  +
  \alpha
  \mathbf{m}
  \times
  \frac{d \mathbf{m}}{dt},
  \label{eq:LLG}
\end{equation}
where $\gamma$ and $\alpha$ are the gyromagnetic ratio and the Gilbert damping constant, respectively. 
In the following, we neglect higher order terms of $\alpha$ 
by assuming that $\alpha \ll 1$, 
which was confirmed experimentally \cite{oogane06,konoto13,iihama14}. 
The magnetic field $\mathbf{H}$ consists of 
an effective perpendicular anisotropy $H_{\rm K}^{\rm eff}$ and the in-plane magnetic field $H_{\rm t}$ as 
\begin{equation}
  \mathbf{H}
  =
  H_{\rm t}
  \mathbf{e}_{x}
  +
  H_{\rm K}^{\rm eff}
  m_{z}
  \mathbf{e}_{z}.
  \label{eq:field}
\end{equation}
The effective anisotropy is defined as $H_{\rm K}^{\rm eff}=H_{\rm K}-4\pi M$, 
where $H_{\rm K}$ is a crystalline anisotropy while $4\pi M$ is a shape anisotropy. 
The magnitude of the applied field is $H_{\rm t}(>0)$. 
The magnetic field is defined as a gradient of energy density $E$ 
as $\mathbf{H}=-\partial E/\partial (M \mathbf{m})$, where 
\begin{equation}
  E
  =
  -MH_{\rm t}
  m_{x}
  -
  \frac{M H_{\rm K}^{\rm eff}}{2}
  m_{z}^{2}.
  \label{eq:energy}
\end{equation}
A spin torque strength $H_{\rm s}$ is given by \cite{slonczewski89,slonczewski05} 
\begin{equation}
  H_{\rm s}
  =
  \frac{\hbar \eta j}{2e (1 + \lambda \mathbf{m}\cdot\mathbf{p}) Md},
  \label{eq:H_s}
\end{equation}
where $d$ is the thickness of the free layer. 
The spin polarization of the pinned layer is $\eta$. 
The parameter $\lambda$ determines the dependence of the spin torque strength 
on the relative angle of the magnetizations, 
and is usually positive \cite{slonczewski89,slonczewski05,xiao04}. 


The self-oscillation can be excited 
when the energy supplied by the spin torque balances 
the dissipation due to damping, 
and thereby, the magnetization precesses on a constant energy curve of $E$. 
This condition can be expressed as $\mathscr{W}_{\rm s}+\mathscr{W}_{\alpha}=0$, 
where $\mathscr{W}_{\rm s}(E)=\gamma M \oint dt H_{\rm s}[\mathbf{p}\cdot\mathbf{H}-(\mathbf{m}\cdot\mathbf{p})(\mathbf{m}\cdot\mathbf{H})]$ 
and $\mathscr{W}_{\alpha}(E)=-\alpha \gamma M \oint dt [\mathbf{H}^{2}-(\mathbf{m}\cdot\mathbf{H})^{2}]$ are 
the work done by spin torque and the dissipation due to damping 
during a precession on the constant energy curve of $E$, respectively. 
This averaging technique over a constant energy curve was developed by several authors 
to investigate the magnetization switching rate in the thermally activated region 
and the oscillation properties in a ferromagnetic multilayer 
\cite{bertotti09book,bertotti04,bertotti06,serpico05,apalkov05,hillebrands06,dykman12,newhall13,taniguchi13PRB1}. 
In general, the magnetization dynamics in the macrospin model is described by two variables. 
The zenith and azimuth angles, 
or the three components in Cartesian coordinate with the constrain $|\mathbf{m}|=1$, 
are often used as the variables. 
Instead, energy $E$ and an angle $\psi$ can be chosen as the two variables, 
where $\psi$ characterizes the angle of the magnetization on a constant energy curve of $E$ \cite{dykman12}. 
The averaging technique over the constant energy curve integrates any quantity over $0 \le \psi \le 2\pi$, 
and reduces the number of the variable to one, $E$. 
The reduction of the number of the variables makes it relatively easy to solve the LLG equation. 
However, there is still a technical difficulty to analytically investigate the possibility to excite self-oscillation 
arising from the calculations of $\mathscr{W}_{\rm s}$ and $\mathscr{W}_{\alpha}$. 
To calculate these functions, the solution of the precession trajectory of the magnetization on a constant energy curve, 
which is described by $d \mathbf{m}/dt = -\gamma \mathbf{m} \times \mathbf{H}$, 
should be derived. 
In the case of a perpendicular field previously studied \cite{taniguchi13}, 
such a trajectory is described by trigonometric functions, 
and the calculations of $\mathscr{W}_{\rm s}$ and $\mathscr{W}_{\alpha}$ become easy. 
On the other hand, in the case of an in-plane field, 
the precession trajectory on the constant energy curve is described by elliptic functions. 
In this case, several mathematical formulas are necessary to derive the explicit forms of $\mathscr{W}_{\rm s}$ and $\mathscr{W}_{\alpha}$. 
The details of the derivation of $\mathscr{W}_{\rm s}$ and $\mathscr{W}_{\alpha}$ 
in the present system are summarized in Appendix. 
Here, we show that the explicit forms of $\mathscr{W}_{\rm s}$ and $\mathscr{W}_{\alpha}$ are given by 
\begin{equation}
\begin{split}
  \mathscr{W}_{\rm s}(E)
  &=
  \frac{\hbar \eta j}{ed}
  \sqrt{
    \frac{r_{1}-r_{3}}{H_{\rm K}^{\rm eff}/(2H_{\rm t})}
  }
  \mathcal{N}_{\rm s},  
  \label{eq:Melnikov_s}
\end{split}
\end{equation}
\begin{equation}
\begin{split}
  \mathscr{W}_{\alpha}(E)
  &=
  -\frac{4\alpha M}{3}
  \sqrt{
    \frac{r_{1}-r_{3}}{H_{\rm K}^{\rm eff}/(2H_{\rm t})}
  }
  \mathcal{H}_{\alpha}. 
  \label{eq:Melnikov_alpha}
\end{split}
\end{equation}
The functions $\mathcal{N}_{\rm s}$ and $\mathcal{H}_{\alpha}$ are given by 
\begin{equation}
\begin{split}
  \mathcal{N}_{\rm s}
  =
  \mp
  &
  \left[
    \frac{(1 \pm \lambda r_{1})}{\lambda^{2}(r_{1}-r_{3})}
    \mathsf{K}(k)
    \pm
    \frac{1}{\lambda}
    \mathsf{E}(k)
  \right.
\\
  &
  \left.
    -
    \frac{1+\lambda^{2} \pm 2 \lambda r_{1}}{\lambda^{2}(r_{1}-r_{3})(1\pm\lambda r_{3})}
    \Pi(k,n)
  \right],
  \label{eq:H_s}
\end{split}
\end{equation}
\begin{equation}
\begin{split}
  \mathcal{H}_{\alpha}
  =&
  H_{\rm t}
  \left(
    \frac{1-r_{1}^{2}}{r_{1}-r_{3}}
  \right)
  \mathsf{K}(k)
  +
  \left(
    \frac{5E}{M}
    +
    3 H_{\rm K}^{\rm eff}
    +
    \frac{2 H_{\rm t}^{2}}{H_{\rm K}^{\rm eff}}
  \right)
  \mathsf{E}(k), 
  \label{eq:H_alpha}
\end{split}
\end{equation}
where $r_{\ell}$ ($\ell=1,2,3$) are given by Eqs. (\ref{eq:r_1}), (\ref{eq:r_2}), and (\ref{eq:r_3}). 
The double sign in Eq. (\ref{eq:H_s}) means 
the upper when $\mathbf{p}$ is parallel to the applied field ($\mathbf{p}=+\mathbf{e}_{x}$) 
and the lower when $\mathbf{p}$ is anti-parallel to the applied field ($\mathbf{p}=-\mathbf{e}_{x}$). 
The first, second, and third kind of complete elliptic integrals are denoted as 
$\mathsf{K}(k)$, $\mathsf{E}(k)$ and $\Pi(k,n)$, respectively, where 
\begin{equation}
  k
  =
  \sqrt{
    \frac{r_{2}-r_{3}}{r_{1}-r_{3}}
  }, 
  \label{eq:modulus}
\end{equation}
\begin{equation}
  n
  =
  \frac{\mp\lambda (r_{2}-r_{3})}{1 \pm \lambda r_{3}}, 
  \label{eq:characteristic}
\end{equation}
are the modulus and characteristic of the elliptic integral, respectively; see Appendix. 
Equation (\ref{eq:Melnikov_s}) is newly derived in this paper, 
while Eq. (\ref{eq:Melnikov_alpha}) was obtained in Ref. \cite{taniguchi15}. 


Equations (\ref{eq:Melnikov_s}) and (\ref{eq:Melnikov_alpha}) are functions of the energy density. 
The range of the energy density is $E_{\rm min} < E < E_{\rm saddle}$, 
where $E_{\rm min}=-(MH_{\rm K}/2)[1 + (H_{\rm t}/H_{\rm K})^{2}]$ 
and $E_{\rm saddle}=-MH_{\rm t}$ are the minimum and saddle point energies, respectively. 
The minimum energy point and the saddle point locate at 
$\mathbf{m}_{\rm min}=(H_{\rm t}/H_{\rm K}^{\rm eff},0,\sqrt{1-(H_{\rm t}/H_{\rm K}^{\rm eff})^{2}})$ 
and $\mathbf{m}_{\rm saddle}=(1,0,0)$. 
The precession period on the constant energy curve of $E$ 
\begin{equation}
  \tau(E)
  =
  \frac{2 \mathsf{K}(k)}{\gamma \sqrt{H_{\rm t} H_{\rm K}^{\rm eff}/2} \sqrt{r_{1}-r_{3}}},
  \label{eq:period}
\end{equation}
relates to the oscillation frequency $f(E)$ via 
\begin{equation}
  f(E)
  =
  \frac{1}{\tau(E)}.
  \label{eq:frequency}
\end{equation}


\section{Analytical Prediction on Possibility to Excite Self-Oscillation}
\label{sec:Analytical Prediction on Possibility to Excite Self-Oscillation}

Let us discuss the possibility to excite the self-oscillation in this system. 
We denote the current density satisfying $\mathscr{W}_{\rm s}(E)+\mathscr{W}_{\alpha}(E)=0$ as $j(E)$, 
\begin{equation}
  j(E)
  =
  \frac{4 \alpha eMd}{3 \hbar \eta}
  \frac{\mathcal{H}_{\alpha}}{\mathcal{N}_{\rm s}}.
  \label{eq:jE}
\end{equation}
The current $j(E)$ in the limit of $E \to E_{\rm min}$ 
is the critical current destabilizing the magnetization in equilibrium, 
which is obtained from the linearized LLG equation as 
\begin{equation}
  j_{\rm c}
  =
  \frac{2 \alpha eMd}{\hbar\eta \mathcal{P}}
  H_{\rm K}^{\rm eff}
  \left[
    1
    -
    \frac{1}{2}
    \left(
      \frac{H_{\rm t}}{H_{\rm K}^{\rm eff}}
    \right)^{2}
  \right],
  \label{eq:jc}
\end{equation}
where $\mathcal{P}$ is 
\begin{equation}
  \mathcal{P}
  =
  \pm
  \left\{
    \frac{(H_{\rm t}/H_{\rm K}^{\rm eff})}{1 \pm \lambda(H_{\rm t}/H_{\rm K}^{\rm eff})}
    \pm
    \frac{\lambda[1 - (H_{\rm t}/H_{\rm K}^{\rm eff})^{2}]}{2 [1 \pm \lambda (H_{\rm t}/H_{\rm K}^{\rm eff})]^{2}}
  \right\}.
  \label{eq:P_eff}
\end{equation}
The double sign in Eq. (\ref{eq:P_eff}) means 
the upper for $\mathbf{p}=+\mathbf{e}_{x}$ and lower for $\mathbf{p}=-\mathbf{e}_{x}$. 
We note here for the later discussion that $j_{\rm c}$ for $\mathbf{p}=-\mathbf{e}_{x}$ changes its sign 
when the applied field magnitude $H_{\rm t}$ becomes 
\begin{equation}
  \widetilde{H}_{\rm t}
  =
  \frac{(1 - \sqrt{1-\lambda^{2}})}{\lambda}
  H_{\rm K}^{\rm eff},
  \label{eq:critical_field}
\end{equation}
whereas $j_{\rm c}$ for $\mathbf{p}=+\mathbf{e}_{x}$ does not. 
As discussed below, the oscillation properties of this type of STO is 
characterized solely by this critical current $I_{\rm c}$. 
This is particularly different from other type of STO 
consisting of an in-plane magnetized free layer, 
where two critical currents are necessary 
to distinguish the in-plane and out-of-plane oscillations \cite{pinna14,newhall13,hillebrands06,taniguchi13PRB1}. 
The number of such critical currents, as well as their values, is determined by the symmetry of the magnetic anisotropy 
and angular dependence of the spin torque, 
and should be investigated for each system. 


\begin{figure}
\centerline{\includegraphics[width=0.8\columnwidth]{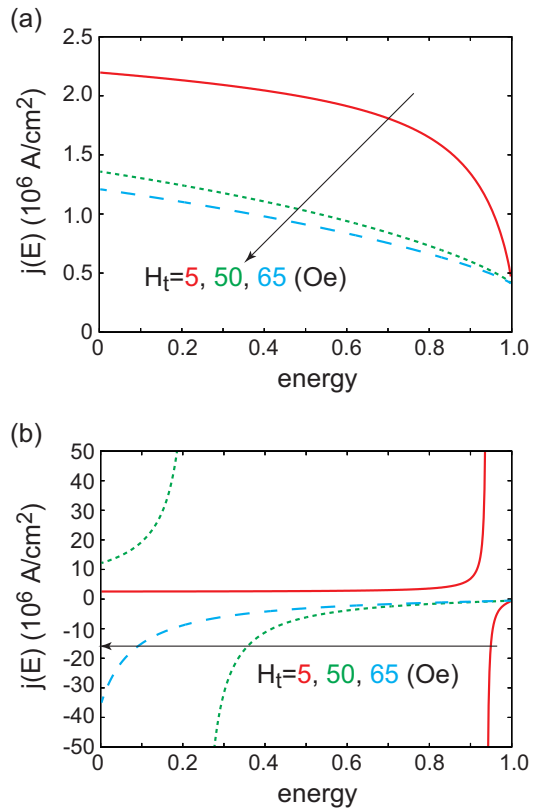}}
\caption{
         The dependence of $j(E)$ on $E$ for (a) $\mathbf{p}=+\mathbf{e}_{x}$ and (b) $\mathbf{p}=-\mathbf{e}_{x}$. 
         The values of the applied field $H_{\rm t}$ are 5 (solid, red), 50 (dotted, green), and 65 (dashed, blue) Oe, respectively. 
         The energy density in the horizontal axis $E$ is normalized as $(E-E_{\rm min})/(E_{\rm saddle}-E_{\rm min})$ 
         to make $E_{\rm min}=0$ and $E_{\rm saddle}=1$. 
         \vspace{-3ex}}
\label{fig:fig2}
\end{figure}



Two conditions should be satisfied to excite self-oscillation on a constant energy curve of $E$ \cite{taniguchi15}. 
The first one is that $j(E)$ is finite. 
More practically, $j(E)$ should be experimentally available value, 
which is at maximum on the order of $10^{7}-10^{8}$ A/cm${}^{2}$ in typical STOs \cite{kiselev03,rippard04,houssameddine07,pribiag07,dussaux10,zeng12,kubota13}. 
The second is that $j(E)/j_{\rm c}>1$. 
This condition relates to the fact that a current larger than $j_{\rm c}$ should be applied to excite 
any kind of magnetization dynamics 
because the magnetization initially stays at the minimum energy state. 
If this condition is not satisfied, 
the magnetization moves to the constant energy curve including $E_{\rm saddle}$, 
and eventually stops its dynamics \cite{taniguchi15}. 
Let us study whether these conditions are satisfied for $\mathbf{p}=\pm\mathbf{e}_{x}$. 
The values of the parameters are taken from the experiments \cite{kubota13}, 
where $M=1448$ emu/c.c., 
$H_{\rm K}=18.6$ kOe, 
$d=2$ nm, 
$\eta=0.54$, 
$\lambda=\eta^{2}$, 
$\gamma=1.732 \times 10^{7}$ rad/(Oe$\cdot$s), 
and $\alpha=0.005$. 
The value of $\tilde{H}_{\rm t}$ is 61 Oe. 
The dependences of $j(E)$ on $E$ for $\mathbf{p}=+\mathbf{e}_{x}$ and $\mathbf{p}=-\mathbf{e}_{x}$ 
are shown in Figs. \ref{fig:fig2} (a) and (b), respectively. 
The values of $H_{\rm t}$ in Fig. \ref{fig:fig2} are chosen as 
$5$, $50$, and $65$ Oe to show $j(E)$ for both $H_{\rm t}<\widetilde{H}_{\rm t}$ and $H_{\rm t}>\widetilde{H}_{\rm t}$.

\begin{figure*}
\centerline{\includegraphics[width=2.0\columnwidth]{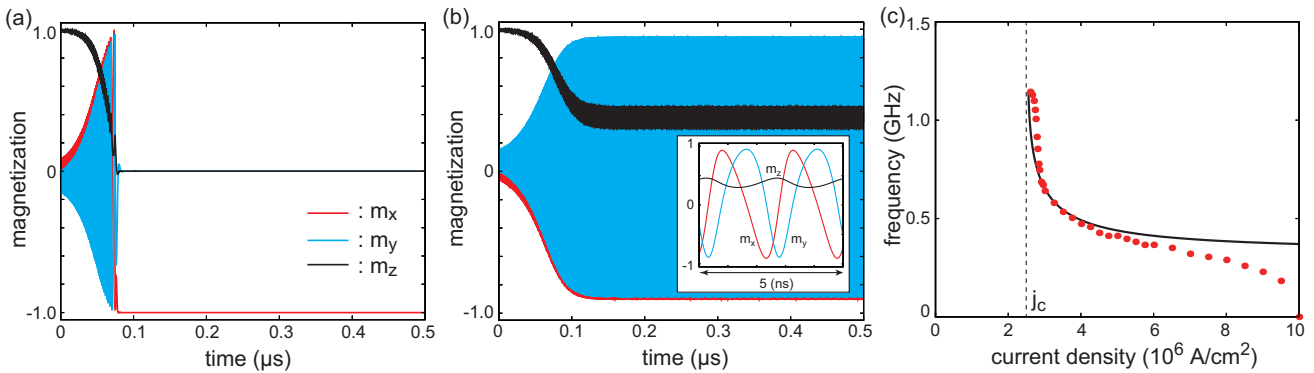}}
\caption{
         The time evolutions of $\mathbf{m}$ at $j=5.0 \times 10^{6}$ A/cm${}^{2}$ with $H_{\rm t}=5$ Oe 
         for (a) $\mathbf{p}=+\mathbf{e}_{x}$ and (b) $\mathbf{p}=-\mathbf{e}_{x}$. 
         The inset in (b) is an enlarged view. 
         (c) The dependence of the oscillation frequency for $\mathbf{p}=-\mathbf{e}_{x}$ on the current density. 
         The dots are obtained by the numerical simulation, while the solid line is obtained from Eqs. (\ref{eq:frequency}) and (\ref{eq:jE}). 
         It should be noted that these equations provide the relation between $f(E)$ and $j(E)$ above $j_{\rm c}$, 
         which is $2.6 \times 10^{6}$ A/cm${}^{2}$ for $H_{\rm t}=5$ Oe. 
         \vspace{-3ex}}
\label{fig:fig3}
\end{figure*}

First, we discuss the possibility to excite self-oscillation 
when the magnetization of the pinned layer is parallel to the applied field ($\mathbf{p}=+\mathbf{e}_{x}$). 
The current $j(E)$ shown in Fig. \ref{fig:fig2} (a) indicates that $j(E)/j_{\rm c}<1$ for $E>E_{\rm min}$, 
meaning that the second condition mentioned above is not satisfied. 
Therefore, a stable oscillation cannot be excited 
when $\mathbf{p}$ is parallel to the applied field. 
This argument will be confirmed in Fig. \ref{fig:fig3} (a) below. 


Next, we discuss the physical interpretation obtained from Fig. \ref{fig:fig2} (b), 
where the magnetization of the pinned layer is anti-parallel to the applied field ($\mathbf{p}=-\mathbf{e}_{x}$).
When $H_{\rm t}<\widetilde{H}_{\rm t}(=61\ {\rm Oe})$, 
the condition $j(E)/j_{\rm c}>1$ is satisfied for the region where $E$ is less than a certain value, 
while $|j(E)|$ diverges at this point. 
For example, $j(E)$ in Fig. \ref{fig:fig2} (b) diverges at 
$(E-E_{\rm min})/(E_{\rm saddle}-E_{\rm min})\simeq 0.94$ for $H_{\rm t}=5$ Oe 
and $(E-E_{\rm min})/(E_{\rm saddle}-E_{\rm min})\simeq 0.23$ for $H_{\rm t}=50$ Oe, respectively. 
These results imply that the self-oscillation can be excited by applying an electric current larger than $|j_{\rm c}|$, 
in which the precession amplitude increases with the increase of the current, 
and finally, saturates to a value corresponding to the energy at the divergence. 
The oscillation frequency given by Eq. (\ref{eq:frequency}) also saturates to a finite value in the large current limit. 
Below, we show that this analytical prediction works well in the low current region, 
whereas a discrepancy between the analytical and numerical results appear in the large current region. 
This discrepancy is due to the breakdown of the assumption used in the averaged LLG equation. 
The divergence of $j(E)$ at a certain $E$ means that 
the net energy supplied by the spin torque during a precession, $\mathscr{W}_{\rm s}$, becomes zero 
when the magnetization arrives at this constant energy curve. 
Then, the spin torque cannot balance the dissipation due to damping even when a large current is applied, 
which means $|j(E)| \to \infty$. 
When $H_{\rm t}>\widetilde{H}_{\rm t}$, the condition $j(E)/j_{\rm c}>1$ is not satisfied, 
and thus the self-oscillation cannot be excited. 


To summarize the above discussion, 
the self-oscillation can be excited 
when the magnetization of the pinned layer and the applied magnetic field points to the opposite directions, 
and the field magnitude is smaller than Eq. (\ref{eq:critical_field}). 
When these conditions are satisfied, 
the analytical theory predicts that 
the oscillation frequency continuously changes from $f(E_{\rm min})$ 
by increasing the current density. 
Equations (\ref{eq:frequency}) and (\ref{eq:jE}) give the relation 
between the current and oscillation frequency. 




\section{Numerical Simulation}
\label{sec:Numerical Simulation}

We confirm the validity of the above discussion 
by comparing with the numerical simulation. 
Figure \ref{fig:fig3} (a) shows an example of the time evolution of $\mathbf{m}$ 
when $\mathbf{p}=+\mathbf{e}_{x}$ 
obtained by numerically solving Eq. (\ref{eq:LLG}). 
The current density, $5.0 \times 10^{6}$ A/cm${}^{2}$, is larger than the critical current density, $2.2 \times 10^{6}$ A/cm${}^{2}$, 
estimated by Eq. (\ref{eq:jc}). 
As shown, the magnetization finally stops at the film-plane, and 
does not show self-oscillation. 
We confirmed similar dynamics when $\mathbf{p}=-\mathbf{e}_{x}$, $H_{\rm t}>\widetilde{H}_{\rm t}$, and $|j|>|j_{\rm c}|$. 
These are consistent with the above discussions. 


On the other hand, 
a stable self-oscillation is excited when $\mathbf{p}=-\mathbf{e}_{x}$ and $H_{\rm t}(=5\ {\rm Oe})<\widetilde{H}_{\rm t}$, 
as shown in Fig. \ref{fig:fig3} (b), 
where the current density is $5.0 \times 10^{6}$ A/cm${}^{2}$ 
while the critical current is $2.6 \times 10^{6}$ A/cm${}^{2}$. 
The dots in Fig. \ref{fig:fig3} (c) are the peak frequencies of the Fourier transformation of $m_{x}$, 
which is measured in experiment \cite{kubota13}. 
The analytical estimation from Eqs. (\ref{eq:frequency}) and (\ref{eq:jE}), shown by the solid line in Fig. \ref{fig:fig3} (c), 
works well in the relatively low current region, 
guaranteeing the validity of the theory. 
On the other hand, a discrepancy between the analytical and numerical results is found in the high current region 
due to the following reason. 
The theory assumes that the averaged work done by spin torque and dissipation due to damping 
during a precession are equal in the self-oscillation state. 
At each point on the constant energy curve however, 
the magnitudes of the spin torque and the damping torque are not equal 
because the spin torque and the damping torque have different angular dependences. 
When the spin torque (current) magnitude or the damping constant is large, 
a shift from the constant energy curve becomes large, 
and thus the magnetization cannot return to the constant energy curve during a precession. 
Then, the assumption in the averaged LLG equation does not stand, 
and the analytical theory is no more applicable to reproduce the numerical results. 
In fact, when the current density is $10 \times 10^{6}$ A/cm${}^{2}$, 
the magnetization does not show self-oscillation. 
In stead, it moves to $\mathbf{e}_{x}$ direction, and stops its dynamics 
because of the large spin torque strength. 
Therefore, the numerically obtained peak frequency in Fig. \ref{fig:fig3} (c) is zero. 


\section{Discussion}
\label{sec:Discussion}

The above analytical and numerical results show that 
a stable self-oscillation can be excited 
when a magnetic field smaller than Eq. (\ref{eq:critical_field}) is applied 
anti-parallel to the magnetization of the pinned layer. 
The oscillation frequency changes continuously by increasing the current magnitude. 
The analytical relation between the current and frequency, 
obtained from Eqs. (\ref{eq:frequency}) and (\ref{eq:jE}), works well 
in the low current region. 
In a high current region, the magnetization eventually arrives at the film-plane, 
and stops its dynamics. 
In this section, we compare these results with the previous works, 
and discuss the advantages to use the in-plane field. 


Let us compare the present results with our previous work \cite{taniguchi13}, 
where the oscillation properties in the presence of a perpendicular field along the $z$-axis were studied. 
It was shown that, when the perpendicular field is smaller than $H_{\rm c}=[3\lambda^{2}/(2-3\lambda^{2})] H_{\rm K}^{\rm eff}$, 
the magnetization discontinuously moves from the perpendicular direction to a large cone angle 
by applying a current larger than the critical current. 
This discontinuous change of the cone angle significantly reduces 
the range of oscillation frequency in the self-oscillation state, as confirmed by experiment \cite{tsunegi14JJAP}. 
Thus, a field larger than $H_{\rm c}$ should be applied to observe 
the oscillation in a wide range of the frequency \cite{kubota13,taniguchi13}. 
On the other hand, in the case of the in-plane field studied here, 
the cone angle of the magnetization continuously changes with the increase of the current. 
Therefore, a wide range of the oscillation frequency can be used. 
Also, the self-oscillation can be excited by a small field less than $\widetilde{H}_{\rm t}$ in Eq. (\ref{eq:critical_field}), 
which is preferable from the view point of a low-power consumption. 
Therefore, the use of an in-plane magnetic field will be an important technical tool for piratical application. 


The system studied above looks similar to an STO with the field-like torque \cite{tulapurkar05,kubota08,sankey08}, 
which was studied in Ref. \cite{taniguchi14APL}, 
because both the field-like torque and 
the torque due to the in-plane field point to $\mathbf{m} \times \mathbf{e}_{x}$ direction. 
However, there are important differences between these torques. 
The strength of the field-like torque depends on the electric current, 
and thus, the spin torque and the field-like torque cannot be manipulated independently. 
On the other hand, the magnitude of the in-plane field 
can be controlled independently from the spin torque. 
The sign of the field-like torque is determined by the material parameters and system size 
\cite{tulapurkar05,kubota08,sankey08,zhang02,zwierzycki05,theodonis06,xiao08}, 
while the direction of the in-plane field is controllable. 
Regarding these points, 
the in-plane field is more convenient than the field-like torque to control self-oscillation, 
although the use of the field-like torque enables to excite self-oscillation 
without an external field. 



Finally, we point out the importance of the parameter $\lambda$. 
The angular dependence of the spin torque, Eq. (\ref{eq:H_s}), characterized by $\lambda$, 
arises from the spin-dependent transport in a ferromagnetic multilayer. 
As can be seen from Eq. (\ref{eq:H_s}), 
the spin torque strength near the anti-parallel alignment of the magnetizations ($\mathbf{m} \simeq -\mathbf{p}$) 
is larger than that near the parallel alignment ($\mathbf{m} \simeq \mathbf{p}$). 
This is because, for example in a GMR system, 
the amount of the spin accumulation, which gives the spin torque on the magnetization, is large 
when the relative angle of the magnetizations is large. 
This parameter had been often neglected in previous works on, for example, 
an in-plane magnetized STO \cite{pinna14} 
because this angular dependence gives small corrections on the oscillation properties such as the critical current. 
However, the parameter $\lambda$ plays the dominant role in STO 
in which the constant energy curve is symmetric with respect to 
an axis perpendicular to the pinned layer magnetization \cite{taniguchi14APEX}, 
as in the case of the present STO with a perpendicular field \cite{taniguchi13}. 
Although the constant energy curve in the this paper is asymmetric due to the in-plane field, 
this parameter still plays a key role to excite the oscillation. 
This is because the field $\tilde{H}_{\rm t}$ satisfies the following relation; 
\begin{equation}
  \lim_{\lambda \to 0}
  \tilde{H}_{\rm t}
  =
  0.
  \label{eq:critical_field_limit}
\end{equation}
This indicates that the self-oscillation cannot be excited by an in-plane field when $\lambda=0$. 
In other words, a finite $\lambda$, which characterizes the asymmetry of the spin torque, 
is necessary to excite the self-oscillation. 
We notice that this conclusion is consistent with the previous study 
on an impossibility to excite self-oscillation by spin Hall effect \cite{taniguchi15}, 
where the spin Hall system corresponds to the limit of $\lambda \to 0$. 



\section{Conclusion}
\label{sec:Conclusion}

In summary, it was shown that 
a stable self-oscillation can be excited 
in an STO with a perpendicularly magnetized free layer and 
an in-plane magnetized pinned layer 
by applying a magnetic field smaller than a critical value, Eq. (\ref{eq:critical_field}), pointing in the direction 
anti-parallel to the magnetization in the pinned layer. 
This is a preferable feature compared with the self-oscillation by a perpendicular field, 
in which relatively large field is necessary to stabilize the self-oscillation. 
The analytical relation between the current and the oscillation frequency was derived by solving the LLG equation. 
The comparison with numerical simulation shows that 
the theory works well in a low current region. 


\section*{Acknowledgement}

The authors acknowledge Takehiko Yorozu for his great help on this work. 
The authors also express gratitude to 
Yoichi Shiota, Satoshi Iba, Hiroki Maehara, and Ai Emura for their kind encouragement. 
T. T. is supported by JSPS KAKENHI Grant-in-Aid for Young Scientists (B) 25790044. 
Y. U. acknowledges JSPS KAKENHI Grants No. 26400390 and No. 26220711. 


\appendix


\section{Derivation of Eqs. (\ref{eq:Melnikov_s}) and (\ref{eq:Melnikov_alpha})}

The functions $\mathscr{W}_{\rm s}(E)$ and $\mathscr{W}_{\alpha}$, 
generally defined as 
\begin{equation}
  \mathscr{W}_{\rm s}(E)
  =
  \gamma M 
  \oint dt 
  H_{\rm s}
  \left[
    \mathbf{p}
    \cdot
    \mathbf{H}
    -
    \left(
      \mathbf{m}
      \cdot
      \mathbf{p}
    \right)
    \left(
      \mathbf{m}
      \cdot
      \mathbf{H}
    \right)
  \right],
  \label{eq:W_s}
\end{equation}
\begin{equation}
  \mathscr{W}_{\alpha}(E)
  =
  -\alpha
  \gamma M 
  \oint 
  dt 
  \left[
    \mathbf{H}^{2}
    -
    \left(
      \mathbf{m}
      \cdot
      \mathbf{H}
    \right)^{2}
  \right],
  \label{eq:W_alpha}
\end{equation}
are the work done by spin torque and the dissipation due to damping 
during a precession on a constant energy curve of $E$. 
The integral is over a precession period, given by Eq. (\ref{eq:period}), 
on the constant energy curve of $E$. 
Thus, the precession trajectory of the magnetization on the constant energy curve 
should be substituted into 
$\mathbf{m}$ in Eqs. (\ref{eq:W_s}) and (\ref{eq:W_alpha}). 
In the present system, such $\mathbf{m}$ is given by \cite{taniguchi15} 
\begin{equation}
  m_{x}(E)
  =
  r_{3}
  +
  (r_{2}-r_{3})
  {\rm sn}^{2}(u,k),
  \label{eq:mx}
\end{equation}
\begin{equation}
  m_{y}(E)
  =
  -(r_{2}-r_{3})
  {\rm sn}(u,k)
  {\rm cn}(u,k)
  \label{eq:my}
\end{equation}
\begin{equation}
  m_{z}(E)
  =
  \sqrt{
    1
    -
    r_{3}^{2}
    -
    (r_{2}^{2}-r_{3}^{2})
    {\rm sn}^{2}(u,k)
  },
  \label{eq:mz}
\end{equation}
where $u=\gamma \sqrt{H_{\rm t} H_{\rm K}^{\rm eff}/2} \sqrt{r_{1}-r_{3}}t$, 
and $r_{\ell}$ are given by 
\begin{equation}
  r_{1}(E)
  =
  -\frac{E}{MH_{\rm t}},
  \label{eq:r_1}
\end{equation}
\begin{equation}
  r_{2}(E)
  =
  \frac{H_{\rm t}}{H_{\rm K}^{\rm eff}}
  +
  \sqrt{
    1
    +
    \left(
      \frac{H_{\rm t}}{H_{\rm K}^{\rm eff}}
    \right)^{2}
    +
    \frac{2E}{MH_{\rm K}^{\rm eff}}
  },
  \label{eq:r_2}
\end{equation}
\begin{equation}
  r_{3}(E)
  =
  \frac{H_{\rm t}}{H_{\rm K}^{\rm eff}}
  -
  \sqrt{
    1
    +
    \left(
      \frac{H_{\rm t}}{H_{\rm K}^{\rm eff}}
    \right)^{2}
    +
    \frac{2E}{MH_{\rm K}^{\rm eff}}
  },
  \label{eq:r_3}
\end{equation}
The Jacobi elliptic functions, 
${\rm sn}(u,k)$, ${\rm cn}(u,k)$, and ${\rm dn}(u,k)$ are useful 
to describe the precession trajectory. 
In the absence of the in-plane field, $H_{\rm t}=0$, 
the system shows the axial symmetry along the $z$-axis. 
In this case, the modulus $k$ becomes zero, 
and thus, the elliptic functions become trigonometric functions. 
Then, $\mathbf{m}$ in Eqs. (\ref{eq:mx}), (\ref{eq:my}), and (\ref{eq:mz}) become 
$m_{x}=-\sin\theta \cos\omega t$, $m_{y}=-\sin\theta \sin\omega t$, and $m_{z}=\cos\theta$, 
where $\theta$ is a constant and $\omega=\gamma H_{\rm K}^{\rm eff}\cos\theta$. 
This solution describes circular motion of the magnetization around the $z$-axis 
with the initial condition $m_{x}(0)=-\sin\theta$. 


Substituting Eqs. (\ref{eq:mx}), (\ref{eq:my}), and (\ref{eq:mz}) into Eqs. (\ref{eq:W_s}) and (\ref{eq:W_alpha}), 
we obtain Eqs. (\ref{eq:Melnikov_s}) and (\ref{eq:Melnikov_alpha}). 
The explicit form of the integral of Eq. (\ref{eq:Melnikov_s}) for $\mathbf{p}=+\mathbf{e}_{x}$ is given by 
\begin{equation}
\begin{split}
  \mathscr{W}_{\rm s}
  =
  \frac{\gamma \hbar \eta j}{2ed}
  &
  \int_{0}^{\tau} 
  dt 
  \left\{
    \frac{H_{\rm t}(1-\lambda r_{3})-H_{\rm K}^{\rm eff}[r_{2}+r_{3}+\lambda(1-r_{3}^{2})]}{\lambda^{2}}
  \right.
\\
  &-
  \frac{H_{\rm t}(r_{2}-r_{3})-H_{\rm K}^{\rm eff}(r_{2}^{2}-r_{3}^{2})}{\lambda}
  {\rm sn}^{2}(u,k)
\\
  &-
  \left.
    \frac{H_{\rm t}(1-\lambda^{2})-H_{\rm K}^{\rm eff}[r_{2}+r_{3}+\lambda(1+r_{2}r_{3})]}
      {\lambda^{2}[1+\lambda r_{3} + \lambda(r_{2}-r_{3}) {\rm sn}^{2}(u,k)]}
  \right\},
  \label{eq:W_s_integral_explicit}
\end{split}
\end{equation}
while that of Eq. (\ref{eq:Melnikov_alpha}) is shown in Ref. \cite{taniguchi15}. 
The explicit form of $\mathscr{W}_{\rm s}$ for $\mathbf{p}=-\mathbf{e}_{x}$ is obtained 
by replacing $\lambda$ with $-\lambda$ and multiplying a minus sign in whole. 
Then, Eqs. (\ref{eq:Melnikov_s}) and (\ref{eq:Melnikov_alpha}) are derived 
by using the following integral formulas \cite{byrd71}; 
\begin{equation}
  \int^{u}
  du^{\prime}
  {\rm sn}^{2}(u^{\prime},k)
  =
  \frac{\mathsf{F}[{\rm am}(u,k),k]-\mathsf{E}[{\rm am}(u,k),k]}{k^{2}},
\end{equation}
\begin{equation}
\begin{split}
  \int^{u}
  du^{\prime}
  {\rm sn}^{4}(u^{\prime},k)
  =&
  \frac{{\rm sn}(u,k) {\rm cn}(u,k) {\rm dn}(u,k)}{3k^{2}}
\\
  &+
  \frac{2+k^{2}}{3k^{4}}
  \mathsf{F}[{\rm am}(u,k),k]
\\
  &-
  \frac{2(1+k^{2})}{3k^{4}}
  \mathsf{E}[{\rm am}(u,k),k],
\end{split}
\end{equation}
\begin{equation}
  \int^{u}
  \frac{du^{\prime}}{b+a{\rm sn}^{2}(u^{\prime},k)}
  =
  \frac{1}{b}
  \Pi[{\rm am}(u,k),k,-a/b],
\end{equation}
where 
\begin{equation}
  \mathsf{F}(\varphi,k)
  =
  \int_{0}^{\varphi} 
  \frac{d \varphi^{\prime}}{\sqrt{1-k^{2}\sin^{2}\varphi^{\prime}}},
\end{equation}
\begin{equation}
  \mathsf{E}(\varphi,k)
  =
  \int_{0}^{\varphi}
  d \varphi^{\prime} 
  \sqrt{ 1 - k^{2} \sin^{2}\varphi^{\prime} },
\end{equation}
\begin{equation}
  \Pi(\varphi,k,n)
  =
  \int_{0}^{\varphi} 
  \frac{d \varphi^{\prime}}{(1-n \sin^{2}\varphi^{\prime})\sqrt{1-k^{2}\sin^{2}\varphi^{\prime}}}.
\end{equation}
are the first, second, and third kind of incomplete elliptic integrals, respectively. 
The Jacobi amplitude function is denoted as ${\rm am}(u,k)$. 
We note that $\mathsf{F}[{\rm am}(u,k),k]=u$, 
$\mathsf{E}[{\rm am}(\mathsf{K}(k),k),k]=\mathsf{E}(k)$, 
and $\Pi[{\rm am}(\mathsf{K}(k),k),k,n]=\Pi(k,n)$, 
where the first, second, and third kind of complete integrals are defined as 
$\mathsf{K}(k)=\mathsf{F}(\pi/2,k)$, $\mathsf{E}(k)=\mathsf{E}(\pi/2,k)$, and $\Pi(k,n)=\Pi(\pi/2,k,n)$, respectively. 




\end{document}